\documentclass[aps,pre,showpacs,amsmath,amssymb]{revtex4}
\usepackage{bm}
\usepackage{graphics}

\begin{document}

\newcommand{\uu}[1]{\underline{#1}}
\newcommand{\pp}[1]{\phantom{#1}}
\newcommand{\be}{\begin{eqnarray}}
\newcommand{\ee}{\end{eqnarray}}
\newcommand{\ve}{\varepsilon}
\newcommand{\vs}{\varsigma}
\newcommand{\Tr}{{\,\rm Tr\,}}
\newcommand{\pol}{{\textstyle\frac{1}{2}}}
\newcommand{\ba}{\begin{array}}
\newcommand{\ea}{\end{array}}
\newcommand{\bea}{\begin{eqnarray}}
\newcommand{\eea}{\end{eqnarray}}
\title{
Two-state dynamics for replicating two-strand systems
}
\author{Diederik Aerts$^1$ and Marek Czachor$^{1,2}$}

\affiliation{
$^1$ Centrum Leo Apostel (CLEA) and Foundations of the Exact Sciences (FUND)\\
Vrije Universiteit Brussel, 1050 Brussels, Belgium\\
$^2$ Katedra Fizyki Teoretycznej i Metod Matematycznych\\
Politechnika Gda\'nska, 80-952 Gda\'nsk, Poland}

\begin{abstract}
We propose a formalism for describing two-strand systems of a DNA type by means of soliton von Neumann equations, and illustrate how it works on a simple example exactly solvably by a Darboux transformation. The main idea behind the construction is the link between solutions of von Neumann equations and entangled states of systems consisting of two subsystems evolving in time in opposite directions. Such a time evolution has analogies in realistic DNA where the polymerazes move on leading and lagging strands in opposite directions.
\end{abstract}
\maketitle

\section{Two-strand systems and mutually time-reflected Turing machines}

According to Adleman \cite{Adleman} the process of DNA replication may be analyzed in terms of Turing machines: One strand plays a role of an instruction tape, a polymeraze is the read/write head, and the second strand contains the results of instructions. 
At a molecular level each strand is a sequence of molecules. In simple models one can represent sequences of molecules in a strand as chains of two-level systems (bits) in a state $|\psi(t)\rangle=\sum_{B_1\dots B_n}\psi(t)_{B_1\dots B_n}|B_1\dots B_n\rangle$, 
$B_j=0,1$. 
Thinking of the motion of the head in terms of a dynamics, one can write 
$|\psi(t)\rangle=U(t,0)|\psi(0)\rangle$, where $0\leq t\leq T$. The final time $T$ is the time of arrival of the head at the end of the strand, and $U(t_1,t_2)$ is a unitary operator which, in principle, may be different for different initial states of the system (this type of evolution occurs for nonlinear Schr\"odinger equations). 

A two-strand system can be represented by an entangled state 
\be
|\Psi(t)\rangle=\sum_{B_j,B_j'}\Psi(t)_{B_1 B_1'\dots B_n B_n'}|B_1\dots B_n\rangle
|B'_1\dots B'_n\rangle.
\ee
The first subtlety we face is that the strands one finds in DNA and RNA are {\it polarized\/}: One side begins with carbon 
$5'$ ($5'$-end), while the other one with carbon $3'$ ($3'$-end) \cite{3-5}. All the natural polymerazes are capable of performing reactions only in the direction $5'\to 3'$. In terms of Turing machines this means that the head always moves in the direction $5'\to 3'$. The second subtlety is that the two strands are anti-parallel: The {\it leading strand\/} begins with $5'$, and the {\it lagging strand\/} begins with $3'$. In consequence, after separation of the two strands, when DNA splits into two Turing machines, the two heads start from opposite ends and move in opposite directions. The dynamics of the entangled state is therefore given rather by $U(t,0)\otimes U(T-t,T)$ than by 
$U(t,0)\otimes U(t,0)$. The two Turing machines are, in this sense, mutually time-reflected. This observation is crucial for the analysis that follows.

\section{Teleological dynamics and two-state formalism}

The evolution operator $U(t,0)\otimes U(T-t,T)$ occurs in quantum mechanics in several contexts, but the one that seems especially relevant here  was introduced by Aharonov, Bergman, and Lebovitz in their analysis of measurements performed in  intermediate times between two other measurements \cite{ABL}. A more modern perspective relates this type of evolution to   quantum systems whose dynamics is constrained by both pre- and post-selection \cite{A1,A2,A3,A4,A5}. 
The post-selection is the important element here: The dynamics one is interested in necessarily ends in a given final state, i.e. is teleological. 

The two states considered by Aharonov and his coworkers form a composite object which is a tensor product of two, in general different, states evolving in time in opposite directions. The state is analogous to a pure-state density matrix, but is in general non-Hermitian and unentangled.

In order to develop more intuitions let us switch for a moment to the $U(t,0)$ given by a nonrelativistic linear 
Schr\"odinger equation. Our two-strand system is given by the wave function 
$
\Psi(t,x,y)=\big( U(t,0)\otimes U(T-t,T)\Psi\big)(x,y)
$
fulfilling 
\be
i\dot \Psi(x,y)=\big(H(x)-H(y)\big)\Psi(x,y)\label{PsivN}
\ee
where $H(x)=-(1/2m)\partial^2/\partial x^2+V(x)$, $H(y)=-(1/2m)\partial^2/\partial y^2+V(y)$.
Let us now introduce the operator $\hat\Psi$ related to $\Psi$ as follows
\be
|\Psi\rangle=\int dx \int dy \Psi(x,y)|x\rangle |y\rangle,\quad
\hat\Psi =\int dx \int dy \Psi(x,y)|x\rangle \langle y|.
\ee
The map $|\Psi\rangle\mapsto \hat\Psi$ is unitary since 
\be
\langle \Phi|\Psi\rangle
&=&
\int dx \int dy\overline{\Phi(x,y)} \Psi(x,y)=\Tr \hat\Phi^{\dag}\hat\Psi
=
(\hat\Phi|\hat\Psi).
\ee
The change of basis $|x\rangle\to U_1|x\rangle$, $|y\rangle\to U_2|y\rangle$ is equivalent to
$|\Psi\rangle\mapsto U_1\otimes U_2 |\Psi\rangle$ and $\hat\Psi\mapsto U_1\hat\Psi U_2^{^\dag}$. 
Eq. (\ref{PsivN}) can be rewritten in a coordinate-free form as the von Neumann equation
$
i d\hat\Psi/dt =[H,\hat\Psi]
$
but the solution $\hat\Psi$ does not have to be Hermitian. 

The interpretation of von Neumann equations in Schr\"odinger terms with the Hamiltonian 
$H\otimes 1-1\otimes H$ (often referred to as a Liouvilian \cite{Bohm}) allows us to define the notions of states of single strands as well as  the distance between the strands. Single strands are defined by the reduced density matrices
\be
{\rho}^\uparrow(x_1,x_2)
&=&
\int dy\,\Psi(x_1,y)\overline{\Psi(x_2,y)},\quad {\rm leading\,strand}\\
{\rho}^\downarrow(y_1,y_2)
&=&
\int dx\,\Psi(x,y_1)\overline{\Psi(x,y_2)},\quad {\rm lagging\,strand}.
\ee
The average distance $D$ between the strands is given by
\be
D^2=
\int dx \int dy (x-y)^2|\Psi(x,y)|^2.
\ee
Since in the present paper we will work with discrete Hilbert spaces of strings of bits, the analogous definitions will read
\be
{\rho}^\uparrow_{j_1j_2}
&=&
\sum_k \Psi_{j_1k}\overline{\Psi_{j_2k}},\\
{\rho}^\downarrow_{k_1k_2}
&=&
\sum_j \Psi_{jk_1}\overline{\Psi_{jk_2}},\\
D^2
&=&
\sum_{j,k}(j-k)^2 |\Psi_{jk}|^2.
\ee
Let us finally note that we can also rewrite the formulas for single-strand states as
\be
{\rho}^\uparrow
&=&
\hat\Psi\hat\Psi{^{\dag}},\\
{\rho}^\downarrow
&=&
\hat\Psi{^T}(\hat\Psi{^T}){^{\dag}}
,
\ee
where $T$ denotes transposition.
\section{Spinorial alphabet}
\label{strands}

Realistic DNA can be labelled either by the four letters ($A$, $C$, $G$, $T$), or the four pairs of letters ($AT$, $TA$, $CG$, $GC$). In both cases we can employ a convention adapted from the two-spinor calculus \cite{PR}. Taking two families of bits, unprimed $B=0,1$ and primed $B'=0',1'$, we can work with the four pairs $(00',01',10',11')$. 
We need one more number, $s=0,1,2,\dots$, to determine the number of strands. 
Following the Introduction we consider 
\be
{\hat \Psi}
&=&
\sum_{B_j,B_j',s,s'}
{\Psi}_{ss';B_1B_1'\dots B_nB_n'}|s,B_1\dots B_n\rangle\langle s',B_1'\dots B_n'|,\\
|\Psi\rangle
&=&
\sum_{B_j,B_j',s,s'}
{\Psi}_{ss';B_1B_1'\dots B_nB_n'}|s,B_1\dots B_n\rangle|s',B_1'\dots B_n'\rangle.
\ee
The number 
\be
\tilde p_{ss';B_1B_1'\dots B_nB_n'}=
|\Psi_{ss';B_1B_1'\dots B_nB_n'}|^2/\parallel \Psi\parallel^2
\ee
represents the probability of finding $s$ leading strands involving the sequence 
$(B_1\dots B_n)$, and $s'$ lagging strands involving the sequence 
$(B'_1\dots B'_n)$. We employ the convention $(AT,CG,GC,TA)=(00',01',10',11')$.

\section{Soliton dynamics of the two strands}

We model the dynamics of two interacting strands by a nonlinear Schr\"odinger equation for $|\Psi\rangle$. We choose the form 
\be
i \dot{\hat \Psi} &=& [H,f({\hat \Psi})].\label{vN}
\ee
where $f({\hat \Psi})$ is, at this stage, unspecified. (\ref{vN}) is a classical Lie-Poisson Hamiltonian system. If the operator $H$ is Hermitian it is convenient to work in the basis of eigenvectors of $H$ since then the diagonal elements ${\Psi}_{jj}$ are time independent. Hamiltonian, Lie-Poisson, or Lie-Nambu structures, as well as the Casimirs associated with (\ref{vN}) were discussed in detail in \cite{PLA97} and generalized in 
\cite{MC-Peyresq}. In particular, for any natural $n$ the quantities $\Tr(H \hat \Psi^n)$ are constants of motion, and  
$\Tr(\hat \Psi^n)$ are Casimir fuctions. 
The link of (\ref{vN}) to generalized thermodynamics was analyzed in 
\cite{MCJN,JNMC}. The fact that (\ref{vN}), with quadratic $f$, is a Darboux-integrable soliton system \cite{MS} was noticed for the first time in \cite{SLMC}, where the dressing-type technique of integration was also introduced. The method was further generalized to a whole hierarchy of von Neumann equations in a series of papers \cite{KCL,UCKL,UC,CCU}. In \cite{Thom} we made a link between such evolutions and general systems involving non-Boolean logic or non-Kolmogorovian probability. Quite recently we showed \cite{I} that the appropriate structures may appear in chemical kinetics and lead to helical configurations.

Now let $\mu, z_\mu,\lambda, z_\lambda$ be complex numbers and consider the two pairs of linear equations for matrices $\varphi_\mu$, $\psi_\lambda$: The ``direct pair"
\be
i\dot\varphi_\mu
&=&
\frac{1}{\mu} f({\hat \Psi})\varphi_\mu\label{LP1'}\\
z_\mu \varphi_\mu
&=&
({\hat \Psi}-\mu H)\varphi_\mu.\label{LP2'}
\ee
and the ``dual pair"
\be
-i\dot\psi_\lambda
&=&
\frac{1}{\lambda} \psi_\lambda f({\hat \Psi})\label{LP1}\\
z_\lambda \psi_\lambda
&=&
\psi_\lambda({\hat \Psi}-\lambda H).\label{LP2}
\ee
The compatibility conditions for the pairs (\ref{LP1'})--(\ref{LP2'}) or 
(\ref{LP1})--(\ref{LP2}) are given by (\ref{vN}). It is essential that the direct and dual pairs are not, in general, mutually Hermitian conjugated since $\hat \Psi$ and $H$ do not have to be Hermitian.

The pairs are Darboux-covariant and were found in \cite{UCKL}.
Both (\ref{LP1})--(\ref{LP2}) and (\ref{vN}) are covariant under the Darboux transformations
\be
\psi_{1,\lambda}
&=&
\psi_\lambda\Big(\bm 1+\frac{\nu-\mu}{\mu-\lambda}\Theta\Big)
\label{DT1}\\
{\hat\Psi}{_1}
&=&
\Big(\bm 1+\frac{\mu-\nu}{\nu}\Theta\Big)
{\hat \Psi}
\Big(\bm 1+\frac{\nu-\mu}{\mu}\Theta\Big)
=
{\hat \Psi}+(\mu-\nu)[\Theta,H]
\label{DT2}
\ee
where $\Theta=\varphi_\mu(\psi_\nu\varphi_\mu)^{-1}\psi_\nu$ and $\psi_\nu$ is any solution  
of (\ref{LP1})--(\ref{LP2}) with $\lambda$ and $z_\lambda$ replaced by new parameters 
$\nu$ and $z_\nu$. A general theory of Darboux transformations in the form we employ here can be found, say, in 
\cite{ZS,NZ,Levi,Mikh,Neu,Cieslinski,LU,LZ,U,L}.
A simple explicit proof of the equivalence of the two forms of ${\hat\Psi}{_1}$ in (\ref{DT2}) can be found in \cite{KCL}.

The matrices occuring in 
(\ref{DT1})--(\ref{DT2}) satisfy $i\dot {\hat\Psi}{_1}=[H,f({\hat\Psi}{_1})]$ and
\be
-i\dot\psi_{1,\lambda}
&=&
\frac{1}{\lambda} \psi_{1,\lambda} f({\hat\Psi}{_1})\label{LP11}\\
z_\lambda \psi_{1,\lambda}
&=&
\psi_{1,\lambda}({\hat\Psi}{_1}-\lambda H),\label{LP22}
\ee
which explains why one speaks of Darboux covariance. This fact can also be used in iterations of the Darboux transformations. 
Let us note that the map ${\hat \Psi}\mapsto{\hat\Psi}{_1}$ switches between different orbits of the same equation. This property can be used for finding nontrivial solutions ${\hat\Psi}{_1}$ on the basis of some known seed solutions ${\hat \Psi}$, in exact analogy to the quantum mechanical method of creation and annihilation operators. Simultaneously, one can employ ${\hat \Psi}\mapsto{\hat\Psi}{_1}$ to model fluctuations between orbits.

\section{Example: Replicating strands}

Let $P_1$ be the projector satisfying $P_1|B\rangle=B|B\rangle$, $B=0,1$, as in  Section~\ref{strands}, and $P_0=I-P_1$ be its orthogonal complement, $P_0|B\rangle=(1-B)|B\rangle$ ($I$ is the $2\times 2$ identity matrix). 
In our present case we take $H=S\otimes \omega N$, where $\omega>0$ will play a role of a kinetic constant, $S$ is the number-of-strands operator, whose eigenvalues are $s=0,1,2,\dots$, and 
\be
N
&=&
\underbrace{P_1\otimes I\otimes\dots\otimes I}_{n}+\dots +
\underbrace{I \otimes\dots\otimes I\otimes P_1}_n.
\ee
The operator $\omega N$ is essentially the Hamiltonian of a system of $n$ non-interacting spins.  
The eigenvalues of $N$ are the natural numbers $0,1,\dots, n$. The spectral representation of $H$ reads 
\be
H &=&\omega\sum_{s=0}^\infty\sum_{m=0}^n s\, m\, \Pi^s\otimes\Pi_m
=
\omega\sum_{s=0}^\infty\sum_{m=0}^n s\,m\, \Pi^s_m,\\
\Pi_m &=& \sum_{B_j=0,1; B_1+\dots+B_n=m}P_{B_1}\otimes\dots\otimes P_{B_n},\\
S &=& \sum_{s=0}^\infty s\Pi^s.
\ee
The degeneracy of the $m$th eigenvalue of $N$ is $n!/(m!(n-m)!)$. One can employ these spectral projectors to turn our von Neumann equation into a formally matrix equation. 

As an example take the nonlinearity $f({\hat \Psi})={\hat \Psi}^q-2{\hat \Psi}^{q-1}$, with any real $q$, solved in a simple case in \cite{UCKL}. 
The eigenvectors of $H$ can be collected into groups corresponding to the same eigenvalue:
\be
|\bm 0^s\rangle
&=&
|s0\dots 0\rangle\nonumber\\
|\bm 1^s_1\rangle
&=&
|s10\dots 0\rangle,\dots,
|\bm 1^s_n\rangle
=
|s0\dots 01\rangle\nonumber\\
|\bm 2^s_{12}\rangle
&=&
|s11\dots 0\rangle,
|\bm 2^s_{13}\rangle
=
|s101\dots 0\rangle,\dots,
|\bm 2^s_{n-1,n}\rangle
=
|s0\dots 11\rangle\nonumber\\
&\vdots&\nonumber\\
|\bm n^s_{12\dots n}\rangle
&=&
|s1\dots 1\rangle\nonumber\\
H |\bm m^s_{j_1j_2\dots}\rangle
&=&
sm\omega |\bm m^s_{j_1j_2\dots}\rangle\nonumber.
\ee
Let us consider the problem of replication of a given sequence of letters. We are interested in the subspace spanned by the vectors 
$|\bm m^0_{j_1j_2\dots}\rangle$ (no strands), $|\bm m^1_{j_1j_2\dots}\rangle$ (one strand), 
$|\bm m^2_{j_1j_2\dots}\rangle$ (two strands). The subspace corresponds to the one spanned by eigenvectors of the Hamiltonian, with the corresponding eigenvalues $0$, $m\omega$, $2m\omega$. 

Generalizing to the present context the strategy from \cite{UCKL} we take the seed solution
\be
\hat{\Psi}_{j_1j_2\dots}(t)
&=&
e^{-iHt}\hat{\Psi}_{j_1j_2\dots}(0)\,e^{iHt}
\\
\hat{\Psi}_{j_1j_2\dots}(0)
&=&
\frac{3}{2}
\Big(|\bm m^0_{j_1j_2\dots}\rangle\langle \bm m^0_{j_1j_2\dots}|
+
|\bm m^2_{j_1j_2\dots}\rangle\langle \bm m^2_{j_1j_2\dots}|\Big)
+
\frac{7}{4}
|\bm m^1_{j_1j_2\dots}\rangle\langle \bm m^1_{j_1j_2\dots}|
\nonumber\\
&\pp=&
-
\frac{1}{2}
\Big(|\bm m^2_{j_1j_2\dots}\rangle\langle \bm m^0_{j_1j_2\dots}|
+
|\bm m^0_{j_1j_2\dots}\rangle\langle \bm m^2_{j_1j_2\dots}|\Big).
\ee
One checks by a straightforward calculation that 
\be
\hat{\Psi}_{j_1j_2\dots}(0)^q-2\hat{\Psi}_{j_1j_2\dots}(0)^{q-1}
&=&
\hat{\Psi}_{j_1j_2\dots}(0)+\Delta_{j_1j_2\dots},
\ee
where
\be
\Delta_{j_1j_2\dots}
&=&
-2\Big(|\bm m^0_{j_1j_2\dots}\rangle\langle \bm m^0_{j_1j_2\dots}|
+
|\bm m^1_{j_1j_2\dots}\rangle\langle \bm m^1_{j_1j_2\dots}|
+|\bm m^2_{j_1j_2\dots}\rangle\langle \bm m^2_{j_1j_2\dots}|\Big)
\nonumber\\
&\pp =&+\frac{1}{4}\Big[1-\Big(\frac{4}{7}\Big)^{1-q}\Big]
|\bm m^1_{j_1j_2\dots}\rangle\langle \bm m^1_{j_1j_2\dots}|,\nonumber\\
{[\Delta_{j_1j_2\dots},H]}
&=&
[\Delta_{j_1j_2\dots},{\Psi}_{j_1j_2\dots}(0)]
=0.
\ee
Analogously, defining 
$
{\hat \Psi}(t) = \sum_{j_1j_2\dots}\hat{\Psi}_{j_1j_2\dots}(t)$,
$\Delta = \sum_{j_1j_2\dots}\Delta_{j_1j_2\dots}$, we find
${\hat \Psi}(t)^q-2{\hat \Psi}(t) = {\hat \Psi}(t)+\Delta$.
In consequence, $[H,{\hat \Psi}(t)^q-2{\hat \Psi}(t)^{q-1}]
=[H,{\hat \Psi}(t)]$ which explains why with this initial condition the solution of 
$i\dot {\hat \Psi}=[H,{\hat \Psi}^q-2{\hat \Psi}^{q-1}]$ coincides with the one of $i\dot {\hat \Psi}=[H,{\hat \Psi}]$. This type of solution of (\ref{vN}) is regarded as trivial. A nontrivial solution is found by means of the Darboux transformation ${\hat \Psi}\mapsto
\hat{\Psi}_{1}$. The solution is positive, i.e. for any projector $P$ we have $\Tr P\hat{\Psi}_{j_1j_2\dots}(t)\geq 0$, but not normalized since 
$\Tr\hat{\Psi}_{j_1j_2\dots}(t)=13/4$. 

The solution of the Lax pair 
\be
z_\nu \psi_\nu(t)
&=&
\psi_\nu(t)\big({\hat \Psi}(t)-\nu H\big),\\
-i\dot\psi_\nu(t)
&=&
\frac{1}{\nu} \psi_\nu(t) f\big({\hat \Psi}(t)\big)
=
\psi_\nu(t) \big(\frac{z_\nu}{\nu}+H+\frac{1}{\nu} \Delta\big)\nonumber
\ee
reads 
$\psi_\nu(t)
=
e^{i\frac{z_\nu}{\nu}t}\psi_\nu(0)
e^{iHt}e^{\frac{i}{\nu}\Delta t}$ with the initial condition satisfying 
\be
z_\nu \psi_\nu(0)
=
\psi_\nu(0)\big({\hat \Psi}(0)-\nu H\big).\ee
Taking $\nu=-i\sqrt{3}/(4m\omega)$
and $\psi_{\nu}(0)
=\langle\phi_{j_1j_2\dots}|=
\frac{1}{\sqrt{2}}\Big(
\langle \bm m^0_{j_1j_2\dots}|
+
e^{-2\pi i/3}\langle \bm m^2_{j_1j_2\dots}|\Big)
$
one verifies that
\be
\langle\phi_{j_1j_2\dots}|
\Big(
\hat{\Psi}_{j_1j_2\dots}(0)-\frac{-i\sqrt{3}}{4m\omega}
2m\omega|\bm m^2_{j_1j_2\dots}\rangle\langle\bm m^2_{j_1j_2\dots}|
\Big)
&=&
\langle\phi_{j_1j_2\dots}|
\Big(
{\hat \Psi}(0)-\frac{-i\sqrt{3}}{4m\omega}H\Big)
\\
&=&
z_{\nu}
\langle\phi_{j_1j_2\dots}|\\
\langle\bm m^1_{j_1j_2\dots}|
\Big(
\hat{\Psi}_{j_1j_2\dots}(0)-\frac{-i\sqrt{3}}{4m\omega} 
m\omega|\bm m^1_{j_1j_2\dots}\rangle\langle\bm m^1_{j_1j_2\dots}|
\Big)
&=&
\langle\bm m^1_{j_1j_2\dots}|
\Big(
{\hat \Psi}(0)-\frac{-i\sqrt{3}}{4m\omega} H\Big)
\\
&=&
z_\nu
\langle\bm m^1_{j_1j_2\dots}|
\ee
with $z_\nu=(7+i\sqrt{3})/4$. 
Accordingly, for any complex $\alpha_{j_1j_2\dots}$, $\beta_{j_1j_2\dots}$, 
the linear combination
\be
\langle\Phi(0)|
&=&
\sum_{j_1j_2\dots}\Big(\alpha_{j_1j_2\dots}
\langle\phi_{j_1j_2\dots}|
+
\beta_{j_1j_2\dots}\langle\bm m^1_{j_1j_2\dots}|\Big)
\ee
satisfies
\be
\langle\Phi(0)|
\Big(
{\hat \Psi}(0)-\frac{-i\sqrt{3}}{4m\omega}H\Big)
&=&
\frac{7+i\sqrt{3}}{4}
\langle\Phi(0)|.
\ee
The solution of the Lax pair and the associated projector $\Theta$ read
($i/\nu=-4m\omega/\sqrt{3}$ is real)
\be
\langle\Phi(t)|
&=&
e^{i\frac{z_\nu}{\nu}t}\langle\Phi(0)|
e^{i Ht}e^{\frac{i}{\nu}\Delta t}\\
\Theta(t)
&=&
\frac{|\Phi(t)\rangle\langle\Phi(t)|}{\langle\Phi(t)|\Phi(t)\rangle}
=
e^{-i Ht}\frac{e^{\frac{i}{\nu}\Delta t}
|\Phi(0)\rangle\langle\Phi(0)|
e^{\frac{i}{\nu}\Delta t}}
{\langle\Phi(0)|e^{2\frac{i}{\nu}\Delta t}|\Phi(0)\rangle}e^{iHt}
\ee
Denoting $\omega_q=m[(4/7)^{1-q}-1]\omega/\sqrt{3}$, 
$|\alpha|^2=\sum_{j_1j_2\dots}|\alpha_{j_1j_2\dots}|^2$, 
$|\beta|^2=\sum_{j_1j_2\dots}|\beta_{j_1j_2\dots}|^2$, 
$|\gamma|=|\beta|/|\alpha|=
e^{-\omega_q t_1}$, 
\be
|\tilde 0\rangle &=&\frac{1}{|\alpha|}\sum_{j_1j_2\dots}\overline{\alpha_{j_1j_2\dots}}|\bm m^0_{j_1j_2\dots}\rangle,\\
|\tilde 1\rangle &=&\frac{1}{|\beta|}\sum_{j_1j_2\dots}\overline{\beta_{j_1j_2\dots}}|\bm m^1_{j_1j_2\dots}\rangle,
\\
|\tilde 2\rangle &=&\frac{1}{|\alpha|}\sum_{j_1j_2\dots}\overline{\alpha_{j_1j_2\dots}}|\bm m^2_{j_1j_2\dots}\rangle,
\ee
and after some calculations we arrive at 
\be
{\hat \Psi}{_1}(t)
&=&
{\hat \Psi}(t)
+
\delta{\hat \Psi}(t)
\\
\delta{\hat \Psi}(t)
&=&
\frac{i\sqrt{3}}{2+2e^{2\omega_q (t-t_1)}}
\Bigg(
\frac{1}{\sqrt{2}}e^{im\omega t}e^{\omega_q (t-t_1)}
|\tilde 0\rangle\langle \tilde 1|
-
\frac{1}{\sqrt{2}}e^{-im\omega t}e^{\omega_q (t-t_1)}
|\tilde 1\rangle\langle \tilde 0|
\\
&+&
e^{-2\pi i/3}e^{2im\omega t}
|\tilde 0\rangle\langle \tilde 2|
-
e^{2\pi i/3}e^{-2im\omega t}
|\tilde 2\rangle\langle \tilde 0|
\\
&-&
\frac{1}{\sqrt{2}}e^{2\pi i/3}
e^{-im\omega t}e^{\omega_q (t-t_1)}
|\tilde 2\rangle\langle \tilde 1|
+
\frac{1}{\sqrt{2}}e^{-2\pi i/3}
e^{im\omega t}e^{\omega_q (t-t_1)}
|\tilde 1\rangle\langle \tilde 2|
\Bigg).
\ee

\section{The simplest case: 1-bit strings}

Let us take a 1-bit system with the Hilbert space spanned by $|s,B\rangle$, i.e.
$
|\bm 0^s\rangle
=
|s,0\rangle
$, $
|\bm 1^s_{1}\rangle
=
|s,1\rangle
$.
The case $m=0$ is trivial since the Hamiltonian $H$ restricted to this subspace vanishes. 
We are thus left with the 3-dimensional subspace spanned by 
\be
|\bm 1^0_{1}\rangle
&=&
|0,1\rangle=|\tilde 0\rangle,\\
|\bm 1^1_{1}\rangle
&=&
|1,1\rangle=|\tilde 1\rangle,\\
|\bm 1^2_{1}\rangle
&=&
|2,1\rangle=|\tilde 2\rangle
\ee
and the Hamiltonian reduces to $H={\rm diag\,}(0,\omega,2\omega)$. 
The matrix 
\be
\hat \Psi_1(t)
&=&
\left(
\begin{array}{ccc}
\frac{3}{2}
& 
i\frac{\sqrt{3}}{4\sqrt{2}}\frac{e^{i\omega t}}{\cosh \omega_q (t-t_1)}
&
\frac{1}{2}e^{2i\omega t}
\Big(-1+i\frac{\sqrt{3} e^{-2\pi i/3}}{1+ e^{2 \omega_q (t-t_1)}}\Big)\\
-i\frac{\sqrt{3}}{4\sqrt{2}}\frac{e^{-i\omega t}}{\cosh \omega_q (t-t_1)}
&
\frac{7}{4}
&
i\frac{\sqrt{3}}{4\sqrt{2}}\frac{e^{i\omega t}}{\cosh \omega_q (t-t_1)}
e^{-2\pi i/3}\\
\frac{1}{2}e^{-2i\omega t}
\Big(-1-i\frac{\sqrt{3} e^{2\pi i/3}}{1+e^{2 \omega_q (t-t_1)}}\Big)
&
-i\frac{\sqrt{3}}{4\sqrt{2}}\frac{e^{-i\omega t}}{\cosh \omega_q (t-t_1)}
e^{2\pi i/3}
&
\frac{3}{2}
\end{array}
\right)
\nonumber\\
&\pp=&
\label{solution}
\ee
solves $id \hat \Psi_1/dt=[H,\hat \Psi_1^q-2\hat \Psi_1^{q-1}]$ for any real $q$. (For small natural $q$ this can be verified directly in Mathematica; for non-natural $q$ one first has to compute the matrix $\hat \Psi_1^{q-1}$ by means of singular value decomposition.) 
The available bits are $B=1$, $B'=1'$ so that the only pair we can find here is $11'=TA$. In Fig.~1 we illustrate properties of the cubic equation ($q=3$) with $m=1$, $\omega=1$, and different values of $t_1=0$, which parametrizes Darboux transformations 
($t_1=\ln (|\beta|/|\alpha|)^{-1/\omega_q}$). The uppermost (solid) curve represents the distance $D$ beween the strands. The distance decreases during the initial phase of replication.
Of course, since one can add to $H$ a multiple of the identity without changing the solution, the matrix (\ref{solution}) solves the equation also for 
$H={\rm diag\,}(k\omega,(k+1)\omega,(k+2)\omega)$, with any $k$. Now the interpretation of the probabilities is different. We have $p_{k,k,TA}$, $p_{k,k+1,TA}$, $p_{k,k+2,TA}$, and so on. 

\section{Example of non-Hermitian $\hat \Psi_1$}

Although the matrix (\ref{solution}) was derived under the assumption that $\omega$ is real, one can check that the solution is still valid if $\omega$ is replaced by a complex parameter
$\omega_c$. This type of continuation of solutions to the complex plane in parameter space has analogies in scattering theory, where real eigenvalues of a Hamiltonian are replaced by complex numbers, and the resulting dynamics describes exponential decay 
\cite{Bohm}. It is interesting that in such a context the complex extension plays an analogous role as in mechanics where complex frequencies are associated with friction forces. Simultaneously, it is known that the decaying states may be obtained by replacing closed-system dynamics by an open-system one. It is therefore tempting to interpret our complex-continuated solutions as describing an open-system dynamics. 
Let us note that the equation  (\ref{vN}) with complex 
$\omega_c=z|\omega|$ and real $t$ is equivalent to the one with real $\omega=|\omega|$ and complex 
$t_c=zt$, and it is known that the map $t\mapsto -it$ switches between Schr\"odinger and heat equations. Still  more subtle reasons for complex ``times" are discussed in \cite{Kaiser}. In our context it is perhaps even more relevant to mention links between replicator equations and matrix-Hamiltonian Schr\"odinger equation with imaginary time \cite{Baake}. 

Anyway, whatever motivation one takes, the resulting $\hat \Psi_1$ is non-Hermitian. Fig.~2 and Fig.~3 illustrate properties of the evolution with, respectively, $\omega_c=1- 0.1 i$ and $\omega_c=-i$, for different values of the parameter $t_1$. The probabilities $p_{ss',TA}$, with $s=s'$, are no longer constant and represent transient effects. The probabilities with $s\neq s'$ evolve irreversibly. Had we replaced $\omega_c$ with their complex conjugated values we would have obtained practically identical plots, only with some colors interchanged. 

The probabilities $p_{k,k,TA}$, $p_{k+1,k+1,TA}$, 
$p_{k+2,k+2,TA}$, $p_{k,k+1,TA}$, $p_{k+1,k,TA}$, become non-negligible only for a finite period of time. The probabilities that stabilize at a non-negligible level (around 1) are 
$p_{k,k+2,TA}$ ($k$ copies of leading 1-bit strands involving $T$ and 
$k+2$ copies of lagging 1-bit strands involving $A$).
The dynamics replaces two $T$s on one strand by two 
$A$s on the other one.

From general properties of the dynamics it follows that the (complex) quantity 
$C=\Tr (H \hat\Psi^2)/\Tr (\hat\Psi^2)$ is a constant of motion. 
However, for non-Hermitian $\hat\Psi$ the quantity 
\be
E^\uparrow
&=&
\frac{(\hat\Psi|H|\hat\Psi)}
{(\hat\Psi|\hat\Psi)}
=
\frac{\Tr (H \hat\Psi\hat\Psi^{\dag})}
{\Tr (\hat\Psi\hat\Psi^{\dag})}
=\frac{\Tr (H \rho^\uparrow)}
{\Tr (\rho^\uparrow)}
\ee
is real but, in general, time dependent. For notational reasons we assume that $H$ is Hermitian (and 
$z$ is incorporated into $t_c=zt$). The function $E^\uparrow$, representing an average energy of the leading strand, is a measure of the energy produced during replication.

\section{Summary and conclusions}

Our analysis started with the following assumptions: 

(1) DNA-type systems consist of pairs of quantum objects (sequences of pairs of molecules) and thus their states should be described in tensor product spaces.
(2)
The dynamics of leading and lagging strands is related by time reversal since the heads of the Turing machines associated with the two strands move in opposite directions. 
(3)
The interaction between the strands is nonlinear and therefore we describe the dynamics by means of a nonlinear Schr\"oedinger equation; the equation is in a one-to-one relation with a Darboux integrable nonlinear von Neumann equation.
(4)
The identification of the tensor structure with a two-strand system allows us to introduce a natural measure of the distance between the strands.
(5)
The Hamiltonian of the model is constructed in a way guaranteeing that values of energy levels are  proportional to the number of strands. 

We have found a particular class of exact solutions and showed that the number of strands is not conserved by the dynamics. Moreover, the change of the number of strands is correlated in time with changes of average distance between the strands. We considered both conservative and dissipative dynamics. The latter was obtained from the former by continuation of solutions to the complex domain in the space of parameters characterizing spectrum of the Hamiltonian. The procedure was motivated by the fact that the continuation was performed within the space of solutions of the dynamical equation, and the very mathematical procedure has analogies in modelling of quantum and classical dissipative systems. 
We have found the solution valid for chains of bits of arbitrary length, but plotted only the simplest 1-bit case. 

The class of integrable von Neumann equations is much wider from what we have explicitly used in this paper. Any soliton von Neumann equation described in \cite{CCU} can be interpreted simultaneously as a helical lattice system, a set of kinetic equations, and a Schr\"odinger type equation for a two-strand system. All these dynamical systems are related to various aspects of DNA-type or similar evolutions. We are yet quite far from the full understanding of the possibilities inherent in integrable von Neumann equations, and their links to chemistry and molecular biology should be further studied. 

\acknowledgments

The work of MC was partly supported by the Polish Ministry of Scientific Research and Information Technology (solicited) project PZB-MIN 008/P03/2003. We acknowledge the support  of the Flemish Fund for Scientific Research (FWO Project No. G.0335.02).

\begin{figure}
\includegraphics{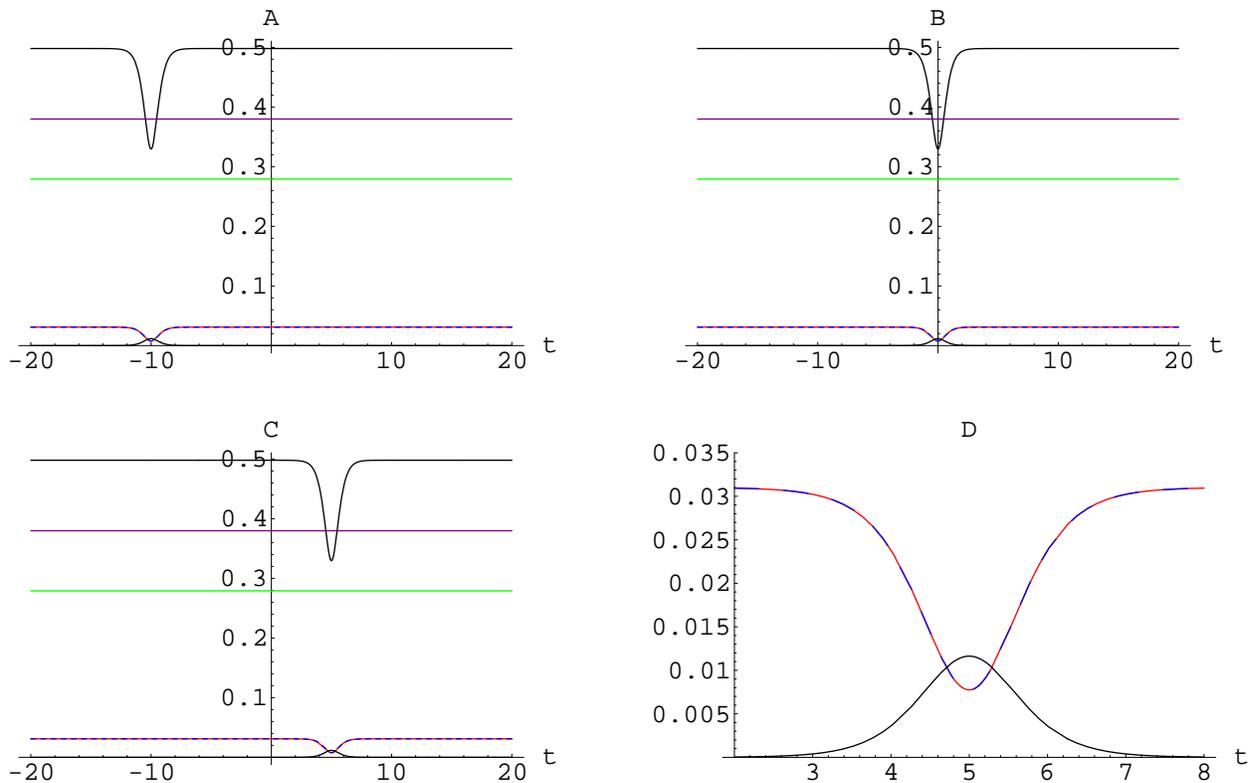}
\caption{The non-dissipative case, $\omega=1$, $m=1$, $q=3$. The parameter $t_1$ controls the effective onset of replication: $t_1=-10$ (A), $t_1=0$ (B), $t_1=5$ (C). Plot (D) is the close-up of plot (C). The uppermost black curve is the distance 
$D(t)$. Probabilities $p_{k+1,k+3,TA}(t)$ (red), 
$p_{k+3,k+1,TA}(t)$ (blue), $p_{k+1,k+1,TA}(t)$ (green), and $p_{k+2,k+2,TA}(t)$ (violet). 
Here red and blue curves overlap, a property that is lost if dissipation is added.}
\end{figure}
\begin{figure}
\includegraphics{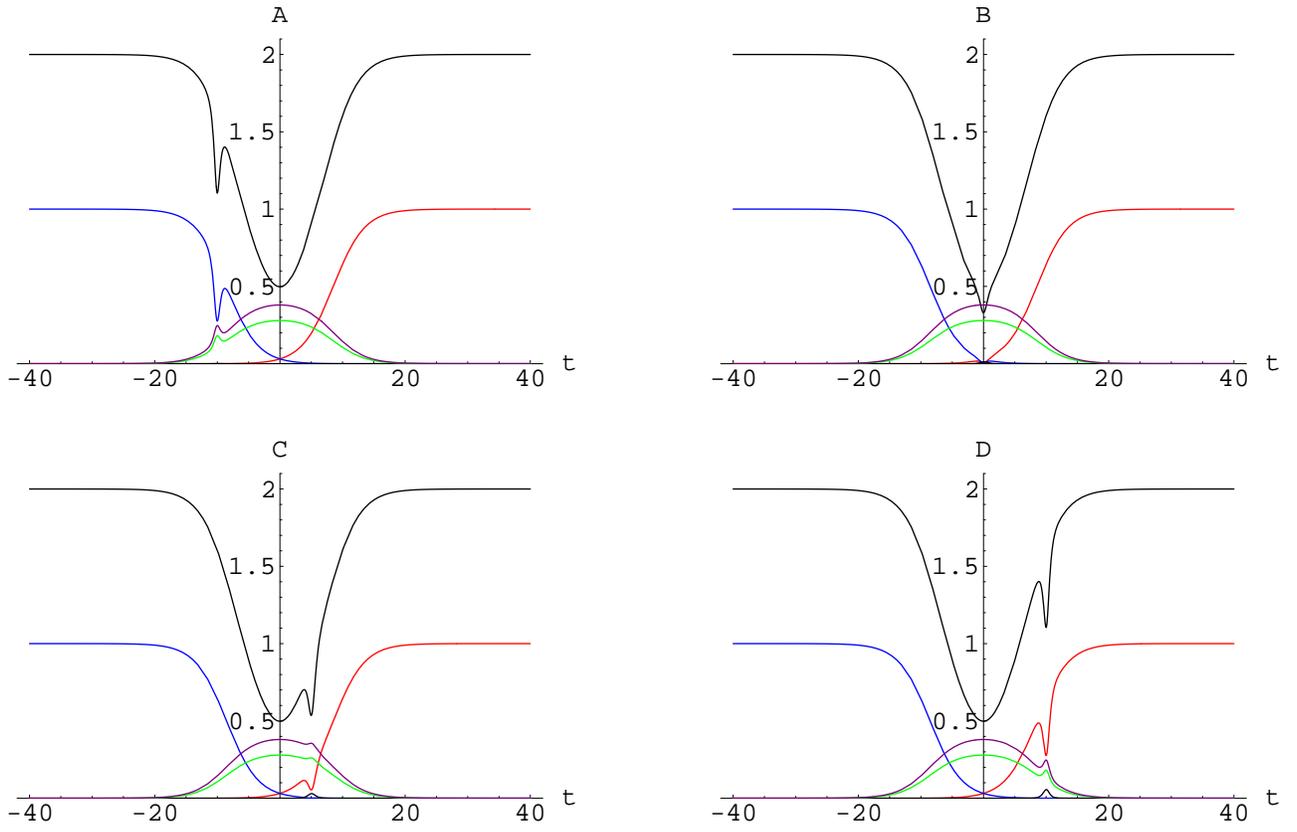}
\caption{The dissipative case with $\omega_c=1-0.1 i$, $m=1$, $q=3$. The position of the dip is controlled by the parameter $t_1$: $t_1=-10$ (A), $t_1=0$ (B), $t_1=5$ (C), 
$t_1=10$ (D). The plot of $p_{k+1,k+2,TA}(t)$ is almost invisible (lowest black curve). }
\end{figure}
\begin{figure}
\includegraphics{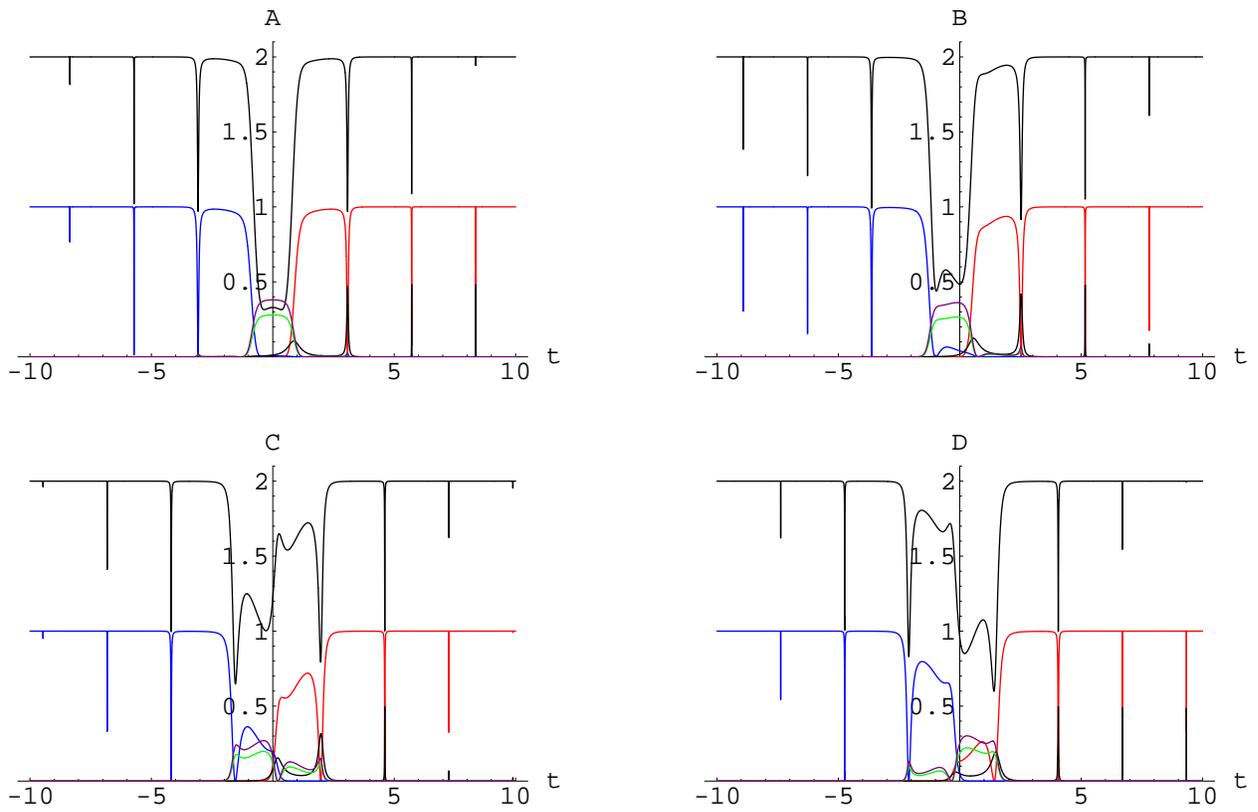}
\caption{The dissipative case with $\omega_c=-i$, $m=1$, $q=3$ (pure diffusion). The notation as in Fig.~2, and $t_1=0$ (A), $t_1=10$ (B), $t_1=20$ (C), 
$t_1=30$ (D).}
\end{figure}

\end{document}